\documentclass[12pt]{iopart}
\usepackage{graphicx}
\usepackage{amsmath}
\usepackage{soul}
\usepackage{color}
\usepackage{booktabs}
\usepackage{xcolor}
\usepackage[backend=biber,style=chem-acs, articletitle=true]{biblatex}
\begin{document}

\title{Thermographic study of freezing water drops: An insight on Mpemba effect}

\author{ Argelia Balbuena Ortega, Emiliano Hernández-Figueroa\\ and J. Antonio del Río\footnote{antonio@unam.mx}}

\address{Instituto de Energías Renovables,\\ Universidad Nacional Autónoma de México,\\ Priv. Xochicalco S/N Temixco, Morelos 62580, Mexico}

%\author{Content \& Services Team}
%\address{IOP Publishing, Temple Circus, Temple Way, Bristol BS1 6HG, UK}
%\ead{submissions@iop.org}
%\vspace{10pt}
%\begin{indented}
%\item[]August 2017
%\end{indented}
%\maketitle

\begin{abstract}\
Despite decades of research, the Mpemba Effect challenges scientists, prompting further investigation and refinement of existing hypotheses. This work uses optical tools such as thermography to analyze and study the Mpemba effect on drops. We analyze times and contact angle changes with temperature with an easily controlled experiment. This work contributes to the ongoing discourse surrounding the Mpemba Effect, emphasizing the need for interdisciplinary collaboration and experimental rigor to unravel the complexities of this intriguing phenomenon. A deeper understanding of the Mpemba Effect enhances our knowledge of thermodynamics and fluid dynamics and opens avenues for practical applications in fields such as cryopreservation, meteorology, and materials science.

%\textbf{Resumen.}
%A pesar de décadas de investigación, el efecto Mpemba permanece desafiante, lo que provoca más investigaciones y refinamiento de las hipótesis existentes. Este trabajo utiliza herramientas ópticas como la termografía para analizar y estudiar el efecto Mpemba sobre gotas de agua. Analizamos tiempos y cambios del ángulo de contacto con la temperatura con un experimento fácilmente controlable. Este trabajo contribuye al discurso actual en torno al efecto Mpemba, enfatizando la necesidad de colaboración interdisciplinaria y rigor experimental para desentrañar las complejidades de este intrigante fenómeno. Una comprensión más profunda del efecto Mpemba mejora nuestro conocimiento de la termodinámica y la dinámica de fluidos y abre vías para aplicaciones prácticas en campos como la criopreservación, la meteorología y la ciencia de materiales.
\end{abstract}

%\maketitle

\section{Introduction}

The Mpemba Effect is a fascinating phenomenon in physics that shows an unusual observation in the cooling process of hot water. Named after Tanzanian student Erasto Mpemba, who noticed it in the 1960s, the Mpemba Effect refers to the counter-intuitive phenomenon where hot water can freeze faster than cold water under certain conditions \cite{jeng2006mpemba}.

The exact mechanisms behind the Mpemba Effect are not entirely understood, and the phenomenon has sparked numerous scientific investigations and debates. Various factors, such as evaporation, dissolved gases, and convection currents, may contribute to the observed effect, making it a complex and intriguing study area \cite{kell1969freezing,wojciechowski1988freezing,esposito2008mpemba,auerbach1995supercooling}. Despite ongoing research, a definitive explanation remains elusive, adding to the mystery and allure of the Mpemba Effect in the realm of thermal physics. Even though it is a well-known phenomenon and has been studied for years, it lacks theoretical models to help interpret experimental results. Only the works by Kell et al. \cite{kell1969freezing} and Vynnycky and Mitchell \cite{vynnycky2010evaporative} provide experimental and theoretical evidence for the effect. 
Parallel to the study of the Mpemba effect, many studies about freezing drops have been done \cite{Chaudhary_Li_2014,Zhang_Wu_Min_2017,Karlsson_Lycksam_Ljung_etal_2019,Yu_Liu_Li_Wang_2023}. The freezing of drops is a multifaceted phenomenon presenting fundamental scientific challenges and practical applications across various fields and industries \cite{ishikawa2004nmr,huang2021overview,zuberi2002heterogeneous}. The Mpemba effect is an example of a false positive in science \cite{trejos}.

There are several questions to be analyzed in freezing  phenomena. One aspect often considered is the rate of cooling. Hot water starts cooling down rapidly when it is in a freezing environment. The rate of heat loss is initially high due to the larger temperature difference between the hot water and the surrounding environment. As the temperature of the hot water approaches that of the cold water, the rate of heat loss decreases.
Another aspect is to analyze the thermal volumetric changes because they depend on the density of water. As the water cools, it generally contracts and becomes denser. However, there is a temperature range (between 0$\,$°C and 4$\,$°C) where water expands as it cools. This unusual behavior is due to the unique structure of water molecules. This ``anomalous expansion'' \cite{Millan_2016} could lead to changes in water molecule distribution and promote ice crystal formation.
Intuitively, this contraction could lead to a rearrangement of the water molecules, potentially promoting the formation of ice crystals more quickly.

The Mpemba effect is counter-intuitive and challenges our fundamental understanding of how heat transfer and freezing work. Investigating the mechanisms behind this phenomenon requires creative thinking, critical data analysis, and a willingness to challenge established assumptions. This phenomenon can be intellectually stimulating and encourage students to think outside the box. The experimental study involves designing and building practical setups, collecting and analyzing data, and developing physical models. Then, in this work, we detailed an experiment of freezing water drops.

Thermography and drops offer a unique lens to explore liquid droplets' thermal signatures and dynamic properties. When applied to droplets, thermography becomes a powerful tool for investigating the heat distribution, cooling, and phase transitions associated with these minuscule entities. 

This study uses traditional and thermographic optical techniques to investigate the Mpemba effect. Using a thermographic camera, we measured the cooling curves and the various temperature changes and processes over time. Meanwhile, with the optical camera, we measure the contact angles at different stages of the cooling process for drops of varying volumes at hot and room temperatures. We use these geometry measurements to analyze the volume change during the cooling process.

The paper's organization is as follows: the next section describes the experimental setup to simultaneously record thermographic and visible videos of the water drop's freezing phenomena. In section \ref{cooling}, we present our results showing the four stages of the cooling phenomena from the hot and environment temperature to complete frozen drops. We emphasize the results of comparing the energy provided for the Peltier cell to the energy used by the phase change phenomenon. We end the paper with a discussion and conclusion on our experiments.

\section{Experimental setup and drop analysis } \label{exp}
In this section, we present the experimental setup and the methodology to analyze the freezing of drops. Also, we present our results by dividing the freezing process into four stages. We focus the analysis on the specific freezing energy and the geometry through the contact angle.

\subsection{Imaging systems}
\begin{figure}
\centering
\includegraphics[width=\linewidth]{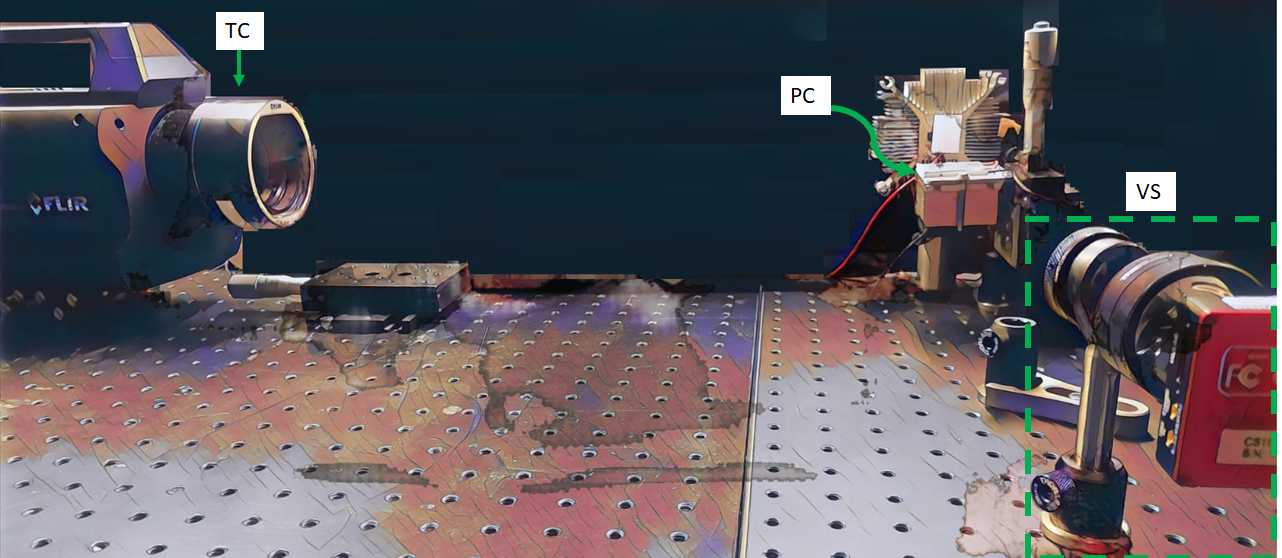}
\caption{Experimental setup: TC is the thermographic camera, PC corresponds to the Peltier cell, and the VS (green box) is the optical visualization system formed by an aperture, a lens with a focal distance of five centimeters, and a CCD camera.}
\label{fig:Setup}
\end{figure}

To investigate the Mpemba effect using thermography, we used the experimental setup shown in Figure \ref{fig:Setup}. The setup comprises two optical imaging systems: a thermographic camera (TC), a visible CCD camera (VS), and a Peltier cell (PC), where the two water drops are deposited to be frozen. The thermographic camera is a FLIR x6540sc with a MWIR 50 mm 1:2.0 USL lens. This device has an Indium Antimonide (InSb) detector, boasting 640 × 512 pixels spatial resolution. The TC operates within a 1.5–5.5$\mu m$ detection range in the mid-wave infrared band and can discern temperature differences as small as 20 mK. The second system (VS) consists of a lens with a focal distance of five centimeters and a CCD Thorlabs camera.

We analyze the freezing process of two drops at different temperatures, which we will call the hot and the room temperature drops. The methodology used for the experiments is as follows: First, we simultaneously started video captures with the thermographic camera and the CCD camera. Seconds after, the hot/room-temperature drop was deposited in the Peltier cell and immediately turned on. We recorded both videos (thermographic and optical) that tracked the cooling process from the initial temperature (hot/room-temperature) to the starting freezing of the drop, then turned-off of the Peltier, and the heating of the drop until its temperature was the same as the room temperature. We repeated this procedure four times for different drop volumes.

In these experiments, we recorded the thermographic videos at 99 frames per second and analyzed them using the ResearchIR 4 software. We traced two indicators in each video to follow the temperature with time, as Figure 2.b illustrates. The first indicator, C, was placed at the drop's center, while the second indicator, P, was placed on the Peltier. The purpose of the Peltier indicator was to have a reference for the actual drop temperature and to know when the measurement would begin since we take the data when the Peltier is on. An example of the typical temperature evolution obtained during each measurement is shown in Figure 2.a. We observe rich information; to describe it, we use the point marked as (a to k). We can observe four behaviors: a monotonic cooling period from a to d. We observe the recalescence from d to e. Drop is freezing from e to h. The drops are frozen from h to i. At i-time, we turn off the Peltier and observe the heating process of the solid drop to the melting point from j to k when the drop reaches the $0\,$ºC at k-time.

We have selected four points in the curve's cooling segments to analyze the energy involved in these processes. The first four points (a-d) are spaced by $1/3$ of the amplitude from the start of the measurement to the recalescence event, while the points from e to h are also spaced by $1/3$ of the maximum temperature from the recalescence to the freezing point. From the cooling curve, we also obtained the amplitude maximum value of the recalescence ($\Delta T_{recal}$) in each measurement.

To observe the freezing process in real-time and measure the contact angles, we used the optical system shown in Figure 2.c. We recorded the standard videos at 33 frames per second, and the contact angle was determined using the ImageJ Drop Analysis plug-in, developed by \cite{Stalder_Kulik_etal_2006}.

\begin{figure}
\centering
\includegraphics[width=\linewidth]{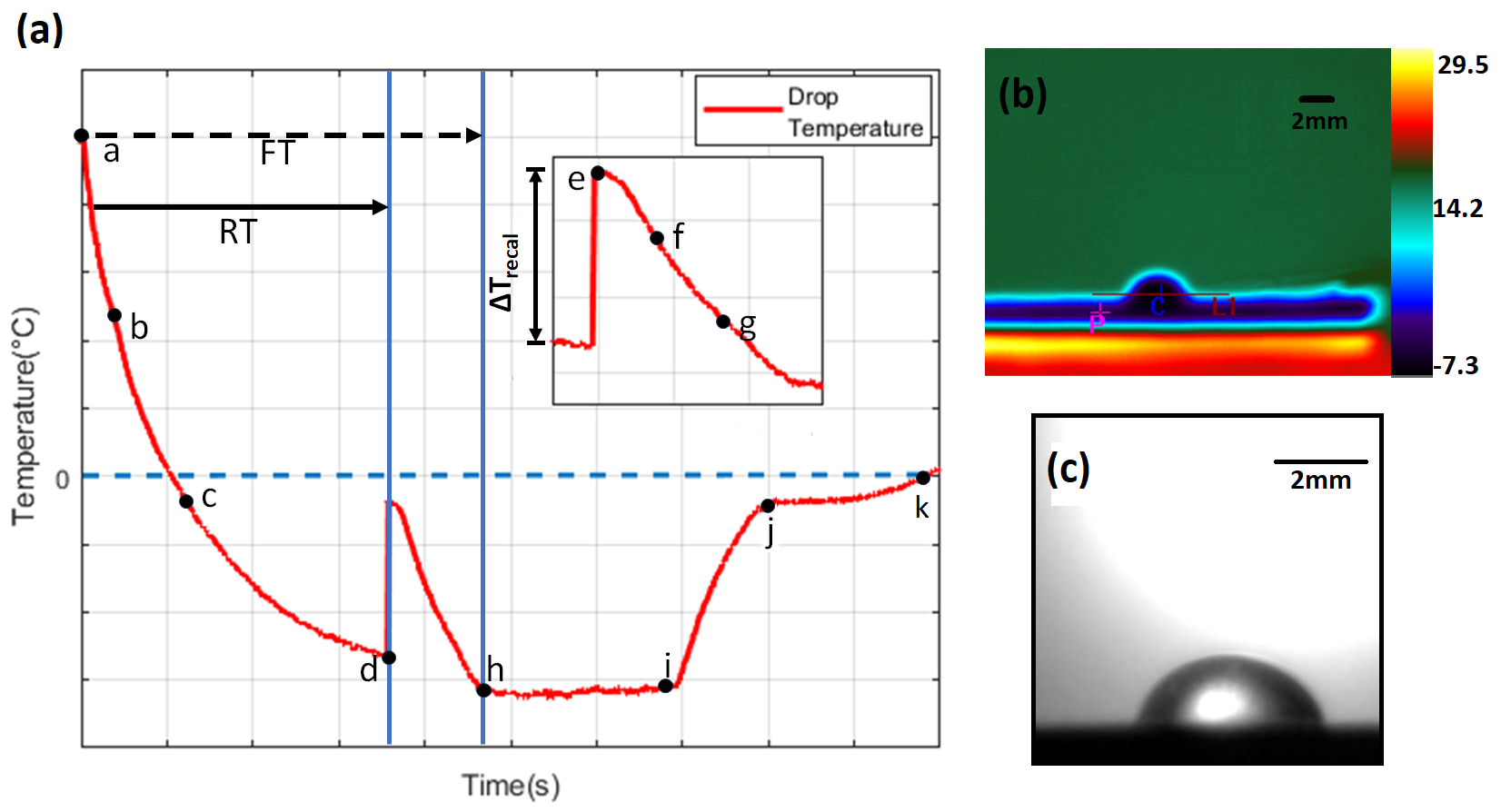}
 \caption{(a) A cooling-heating temperature graph was obtained with the FLIR image software. The points marked from a to h are the selected moments of the droplet freezing process. The solid arrow shows the time from the start of cooling to the recalescence (RT), while the dotted arrow shows the time from the cooling start to the drop freezing (FT). $\Delta T_{recal}$ is the temperature variation caused by the recalescence. (b) Infrared images, C and P, are the indicators for obtaining the cooling curves. (c) Example of visualization of the drop and the measure of the contact angles}
    \label{fig:Parametros}
\end{figure}

We need more detailed explanations of some of these different behaviors. While the cooling stage (a-d) is straightforwardly understood, the recalescence (e-h) stage is not always analyzed or described. Recalescence is a physical phenomenon that occurs during the freezing of water droplets. It is the sudden temperature rise that occurs when a supercooled water droplet freezes due to the release of latent heat. During the recalescence stage, the water droplet becomes opaque  \cite{jingru}. Also, the energy transfer involved in the freezing process requires description. 
In the following section, we present our results for nine different drop volumes ($1$, $3.5$, $5$, $7.5$, $10$, $12.5$, $15$, $17.5$ and $20 \mu L$) for room and hot temperature distilled water.

\section{Cooling process: temperature and geometric features}\label{cooling}
%\subsection{Results}

This section presents the results in two physical aspects: the cooling process on time and contact angle behavior. These two aspects give insights into understanding the physical phenomena needed to observe the Mpemba effect.

\subsubsection{Cooling, recalescence and freezing processes}

\begin{figure}
    \centering
    \includegraphics[width=\linewidth]{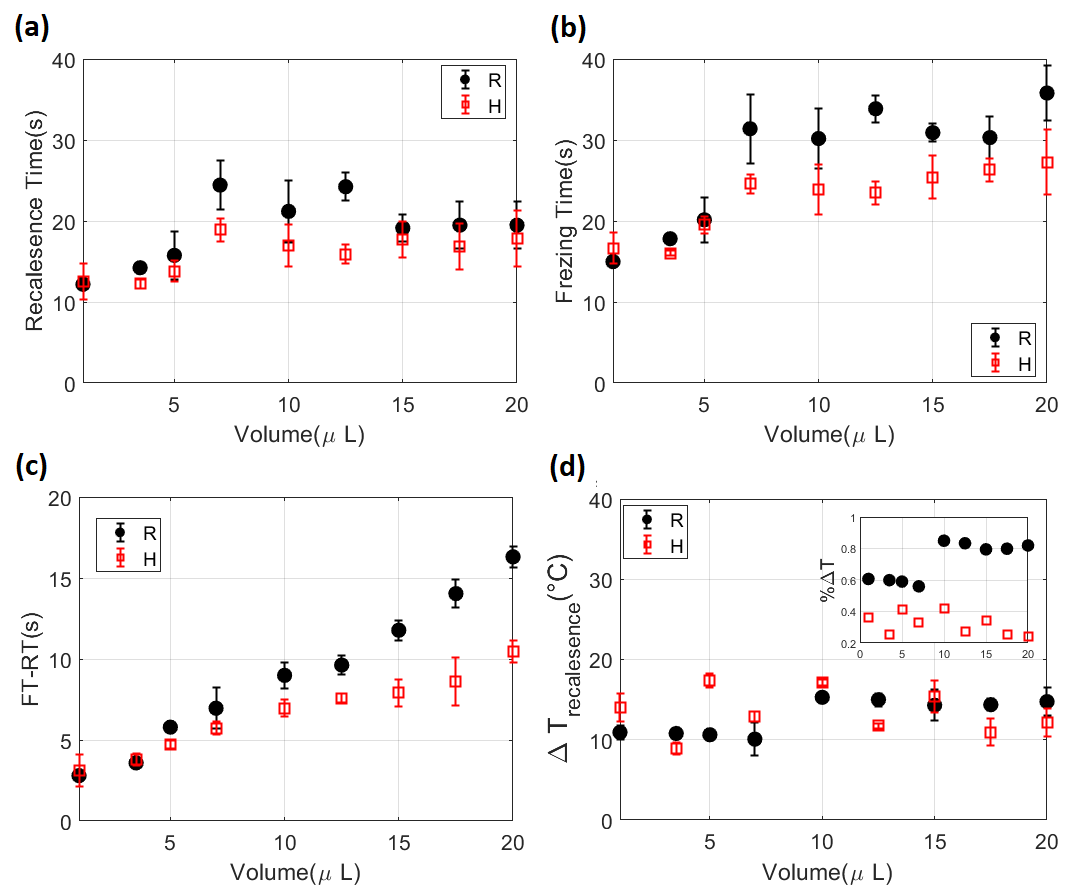}
    \caption{The following data has been collected regarding the freezing process of water drops as a function of their volume. (a) The time from the beginning of cooling to the point of recalescence (RT). (b) The time from the beginning of cooling to the point of drop freezing (FT). (c) The time from recalescence to drop freezing (FT-RT). (d) The temperature amplitude during recalescence, with the inset showing the percentage related to the maximum temperature achieved in each measurement. In all the graphs, the filled circles represent data of drops at room temperature, while the empty squares represent results obtained for hot drops.}
    \label{fig:DatosTiempo}
\end{figure}

Figure \ref{fig:DatosTiempo} shows the relevant findings which characterize the cooling process. In all the graphs, the filled circles represent data of drops at room temperature, while the empty squares represent values obtained from hot drops. 

As we mentioned, there are four stages of droplet freezing based on their temperature: liquid-cooling, supercooling, freezing, and solid-cooling. Other authors categorize them into two general stages: nucleation/supercooling and solidification \cite{Zhang_Liu_Min_Wu_2019}. The statement suggesting ``hot water can freeze faster than cold water'' can be ambiguous without proper consideration. Then, it is crucial to clarify whether this refers to the time it takes for ice to begin forming or the time needed for the entire sample to freeze \cite{auerbach1995supercooling}. With the latter consideration in mind, the first time we measured was RT time. After it, recalescence occurs, and the falling temperature rises momentarily before stabilizing. This increase in temperature during the cooling process is understood because the solidifying material releases latent heat, which can cause a brief temperature increase \cite{Chaudhary_Li_2014, Yu_Liu_Li_Wang_2023}. As a result of recalescence, the droplet transforms from a supercooled liquid to a mixture of water and ice with uniform temperature, the equilibrium temperature. During freezing, the phase change propagates from the droplet base while the rest remains at equilibrium temperature. Ice nucleation of a supercooled droplet has randomness, and its occurrence requires a significant degree of supercooling \cite{Zhang_Wu_Min_2017}. Figure \ref{fig:DatosTiempo}.a shows this time (RT) for the different volumes; we observed no clear tendency of volume dependence with this time. However, the hot drops tend to reach point d faster than the room temperature drops.

The second period we analyzed was FT (Figure \ref{fig:DatosTiempo}.b), which includes the actual freezing stage. In this period we can observe the Mpemba effect since it is evident the difference in the time it takes, on average, for the hot drops to freeze compared to those at room temperature. However, this observation is clearer for large volumes. In contrast, it might be negligible for small volumes (1 to 5 µL) because the drop's mass is tiny and they have no significant thermal inertia. When the drop hits the Peltier, its temperature decreases relatively fast, especially if it is a hot droplet. Thus, the difference between experiments is minimal. It was noticed that a temperature difference $>=$ 20 °C between the hot and the room droplet is needed in our system for the Mpemba effect to be notable. Also, the experimental results show that the FT grows slowly with the volume.

We measured the time it takes from the recalescence to the total solidification of the drop (FT-RT), which is generally the time reported in most Mpemba effect studies. As shown in Figure \ref{fig:DatosTiempo}.c, there is a dependence between the volume and this time for both cases (hot/room temperature). However, the increase in time with respect to volume is more pronounced for drops at room temperature than for hot drops.

Finally, we analyze the energy involved in the complete cooling process. Since the power of the Peltier cell is a constant $P_{p}=25.83 W$, these values of time can be associated with the amount of energy necessary to freeze the drop  $E_{f}=P\cdot(FT-RT)$. Directly from this relationship, we can infer that the energy required to freeze a drop at room temperature is greater than that required to freeze a drop at hot temperature and that there is a dependence on the volume.

Another aspect that caught our attention was the amplitude of the recalescence; since this is an effect of latent heat, we expected it to be related to the volume of the drop, but we found that it is not linearly related. However, the volume plays a role in it, Figure \ref{fig:DatosTiempo}.d shows this relation. The value of the recalescence is, on average, $12.5\,$ºC for the room temperature drops and $13.22\,$ºC for the hot drops. The inset shows the percentage relative to the maximum temperature achieved in each measurement.
%; nevertheless, it seems that an increment of volume will give; as a result, an increment of this value.  

\subsubsection{Specific freezing energy}

We observed water subcooling before the recalescence and solidification stage for all drop-freezing tests. It was proposed, in contrast to the phenomenon of change of state of water at $0\,$ºC, that the specific energy required to achieve drop freezing is larger than simply the product of the droplet mass and the latent heat since the entire droplet changes to the solid state at temperatures well below the freezing point.

Equation \eqref{eq:E-f-sum} defines the specific total freezing energy $E_{ft}$ as the sum of the latent heat of freezing of water and three specific energies: the energy yielded from the time the drop reaches $0\,$ºC to the instant before nucleation ($E_{0-d}$), the energy change present at recalescence ($E_{d-e}$), and the energy yielded from the end of recalescence to total solidification ($E_{e-h}$). The specific energies were calculated by following the expressions \eqref{eq:E-0-d}, \eqref{eq:E-d-e} and \eqref{eq:E-e-h} where $c_p = 4.22 \frac{kJ}{kg\cdot K}$ is the specific heat of water at 0 ºC and $L = -333.7 \frac{kJ}{kg}$ is latent heat for freezing.
\begin{equation}
    E_{ft} = E_{0-d} + E_{d-e} + E_{e-h} + L
    \label{eq:E-f-sum}
\end{equation}
\begin{subequations}
    \begin{align}
        E_{0-d} = c_p \cdot \Delta_{0d} \label{eq:E-0-d}\\
        E_{d-e} = c_p \cdot \Delta_{de} \label{eq:E-d-e}\\
        E_{e-h} = c_p \cdot \Delta_{eh} \label{eq:E-e-h}
    \end{align}
\end{subequations}

\begin{figure}
    \centering
    \includegraphics[width=\linewidth]{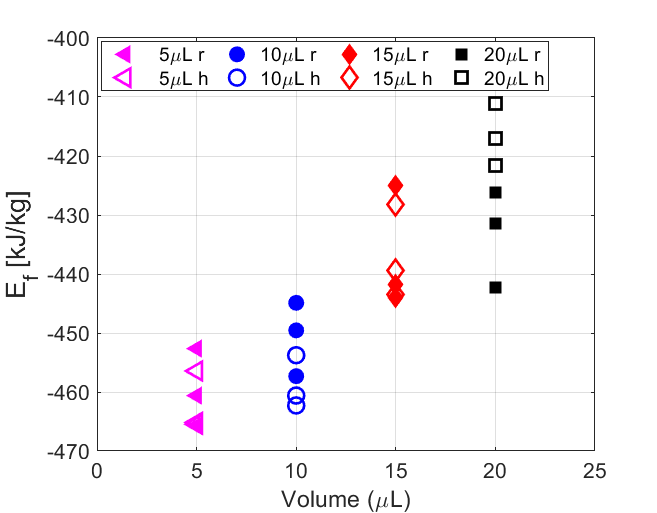}
    \caption{Specific freezing energy for different drop volumes for hot and cold drops.The filled circles represent data of drops at room temperature, while the empty squares represent results obtained for hot drops}
    \label{fig:E-f}
\end{figure}

Where $\Delta_{0d}=T_d-T_0$ is the difference of drop temperature from $0^\circ$ to the point d, $\Delta_{de}=T_e-T_d$ is the difference of temperature from the recalescence and the point d, and $\Delta_{eh}=T_e-T_h$ is the difference of temperature from the recalescence to the complete freezing.
The total specific freezing energy was calculated for 5, 10, 15, and 20 droplets $\mu L$. The results, shown in figure \ref{fig:E-f}, clearly depend on volume. In other words, the energy per unit mass removed from the droplet by the Peltier cell to achieve solidification is higher for droplets of smaller volume than for droplets of larger volume.
From the results, there is no apparent difference between the hot and cold drops when the volumes are small; however, the higher amount of energy for the room temperature drop to totally freeze is more evident when the volume of the drop increases. 

\subsubsection{Geometric changes: Contact angles}

We measured the left and right contact angles of each drop at specific times (a to h, as shown in Figure \ref{fig:Parametros}.a) using the Drop Analysis plug-in for ImageJ \cite{Stalder_Kulik_etal_2006}. The plug-in allowed us to save the coordinates of the knots used to find the contact angles of the drop. We saved the knots for each drop and each time in .txt files and processed them in MATLAB. In the literature, there are many forms to measure the drops' contact angle \cite{faruk}.  

\begin{figure}
    \centering
    \includegraphics[width=\linewidth]{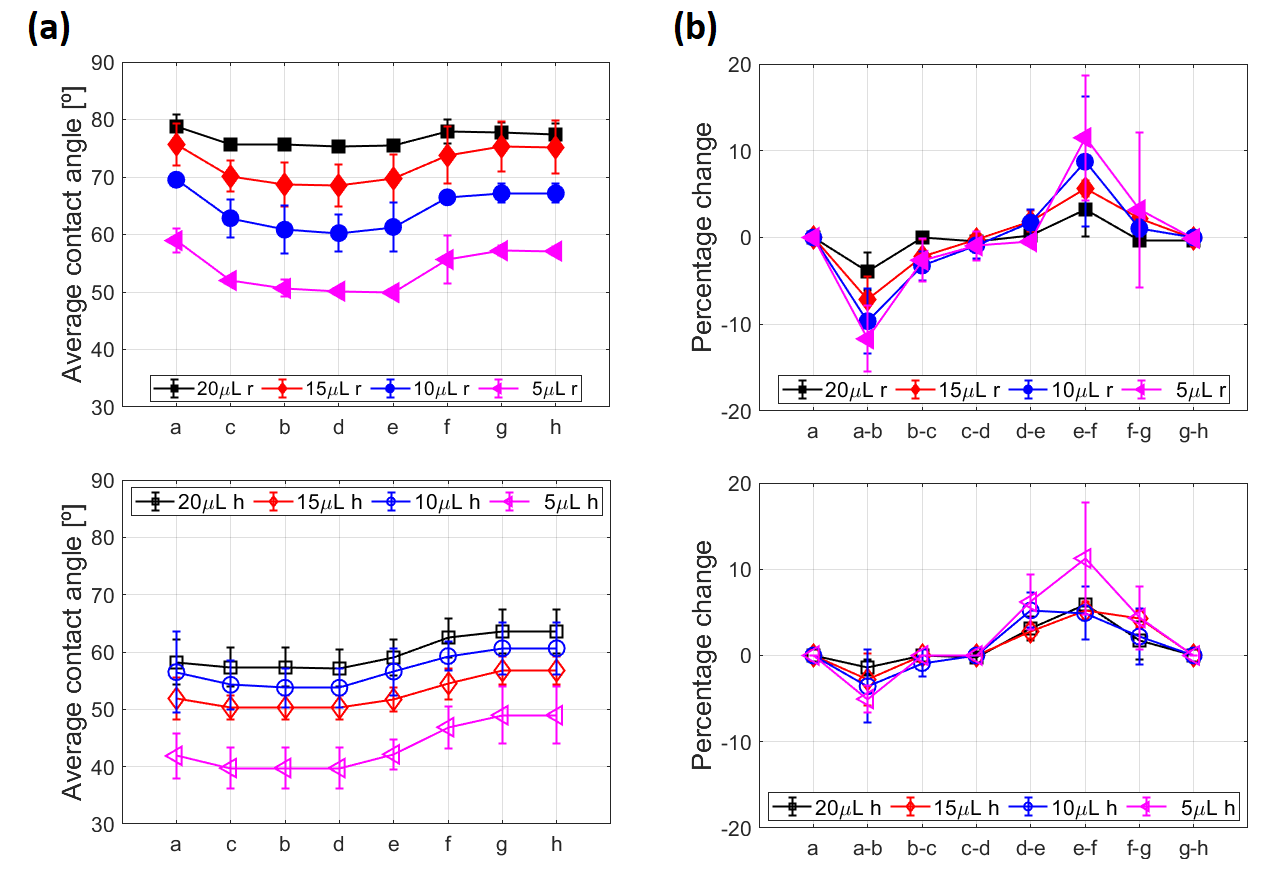}
    \caption{(a)The average contact angles for the drops at room temperature (first row ) and the drops with hot water (second row).(b) Percentage difference in the contact angle between the different selected points (see figure \ref{fig:Parametros}), the first row corresponds to the drops at room temperature while the second corresponds to the drops with hot water.
 }
    \label{fig:AC}
\end{figure}

Figure \ref{fig:AC}.a displays the average contact angles for drops at room and hot temperatures (first and second row, respectively) for four selected volumes. Due to the difference in water density, there was a $25\%$ initial variation in the contact angle (at point a) between the room temperature and hot temperature drops. We observed a decrease in the contact angles for both temperatures (from b to e), a return to their initial values in the case of room temperature drops, and an increase of $8\%$ for hot drops (at point h). To better visualize these changes, we plotted the percentage difference in contact angles between the different selected points (a to h) in Figure \ref{fig:AC}.b. The most significant changes occurred between a and b, and e and f during the cooling process. These corresponded to the first third of the cooling and from the recalescence maximum to the first third of the nucleation. We also observed that the changes were minor for the hot droplets compared to the room-temperature ones. This difference indicates that the energy required to freeze the drops at room temperature is higher than in the case of hot droplets. Therefore, if we have a Peltier with constant power, the time required to freeze the room-temperature drops should also be higher. This observation indicates the Mpemba effect.

\begin{figure}
    \centering
    \includegraphics[width=0.8\linewidth]{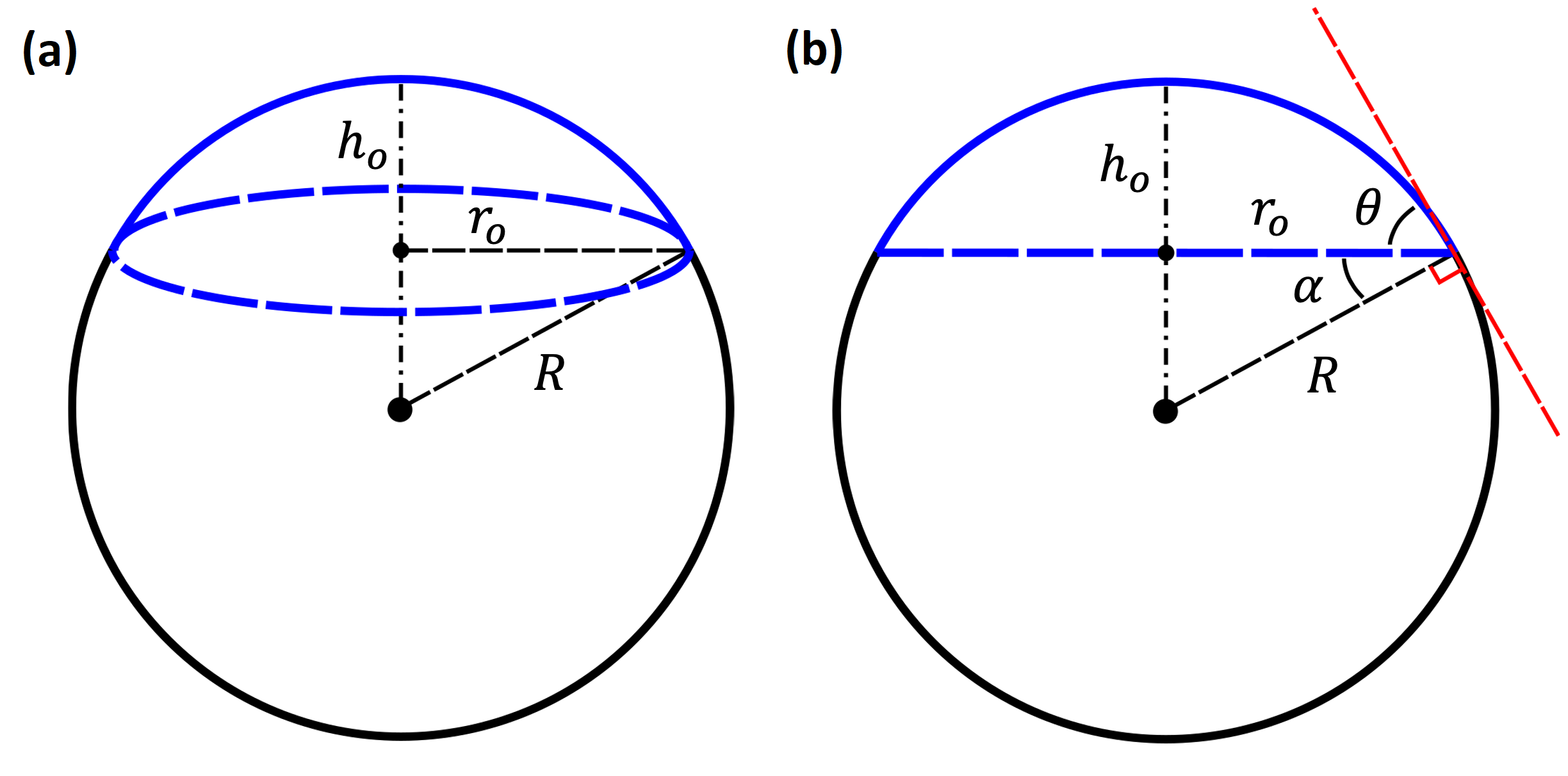}
    \caption{Graphical scheme for calculating the drop volume depending on the contact angle. (a) Three-dimensional view of a spherical cap with radius $R$, (b) Two dimension scheme: $\theta$ is the contact angle measured,  $r_{0}$ is the distance between the edge of the drop and the center of it,  $h_{0}$ is the high of the drop and $\alpha$ is the alter intern angle.}
    \label{fig:CV}
\end{figure}

On the other hand, we noticed a change in the drop volume in each experiment; to verify this, we calculated it through the contact angles. 
Based on its contact angles, the drop geometry was assumed to be a spherical cap to estimate its volume. This simplification does not consider gravity effects and is valid because the contact angles are less than $90^\circ$. This approximation would not be appropriate for hydrophobic surfaces with a higher contact angle. Figure \ref{fig:CV}.a  schematizes the spherical cap model of radius $R$, while  Figure \ref{fig:CV}.b presents a 2D visualization closest to the experimental image.

\begin{figure}
    \centering
    \includegraphics[width=\linewidth]{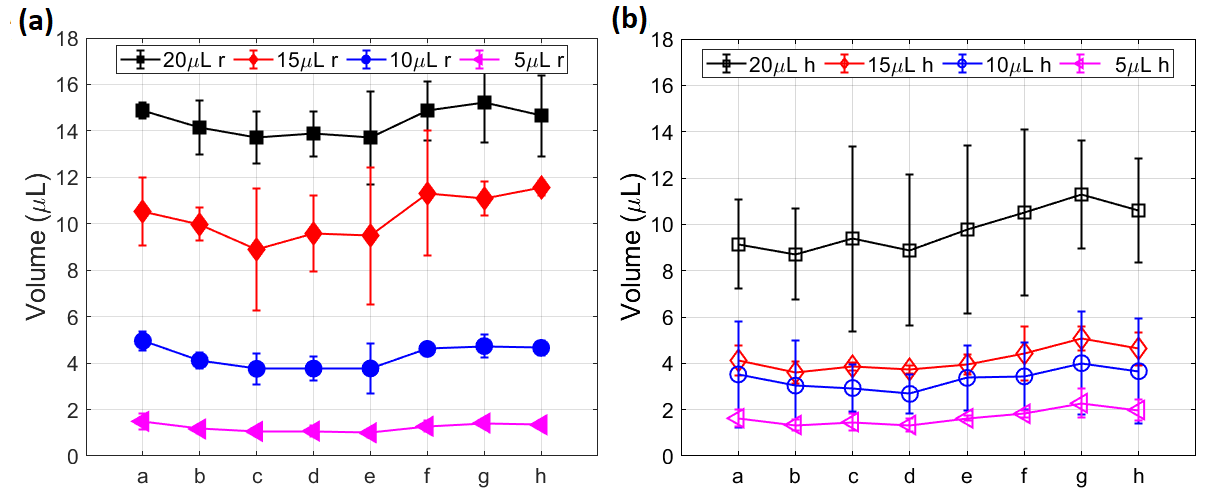}
    \caption{Calculated volume depending on the contact angle for the (a) Room temperature drops and for the (b) hot drops.}
    \label{fig:Vol}
\end{figure}

Then, the volume of a spherical cap \cite{Polyanin_Manzhirov_2006} can be calculated using the following expression:
\begin{equation}
V=\frac{\pi}{6}h_0(3r_o^2+h_o^2)
\label{eq:v}
\end{equation}
where  $r_{0}$ is the distance from the center of the cap to the edge of it, and $h_{0}$ represents the drop height. Through trigonometric relationships, for angles less than $90^\circ$, $h_o$ can be expressed as a function of the contact angle $\theta$ and the radius of the spherical cap, $R$, as shown in equation \ref{eq:ho}.
\begin{equation}
ho = R (1-sin(90^\circ-\theta)) 
\label{eq:ho}
\end{equation}
Combining both equations \eqref{eq:v} and \eqref{eq:ho}, we obtained the volume of the drop for the selected times (a to h); $r_o$ and $R$ were obtained with a MATLAB script, fitting a circular geometry to the known knots of the drop, previously acquired. Figures \ref{fig:Vol}.a and \ref{fig:Vol}.b show the calculated volume depending on the contact angle for the room temperature drops and hot drops. Notice that in both graphs, the volume diminishes from the start of the cooling to the recalescence, consistent with the variation of the contact angles. 

Verification was carried out in order to gain better insight into the variation of the calculated volume. This verification consisted of multiplying the volume by the change in temperature from one measurement moment to another and dividing it by the time between measurements, as follows: 

\begin{equation}
\frac{V(\theta)\Delta T}{\Delta t}=\frac{P_{PC}}{\rho c_d}
\label{eq:VT}
\end{equation}
where $P_{PC}$ is the power of the Peltier cell, the $c_d$ is the specific heat of the water drop at an arbitrary moment of cooling/freezing, and $\rho$ is the density of water, which depends on temperature.

\begin{figure}
    \centering
    \includegraphics[width=\linewidth]{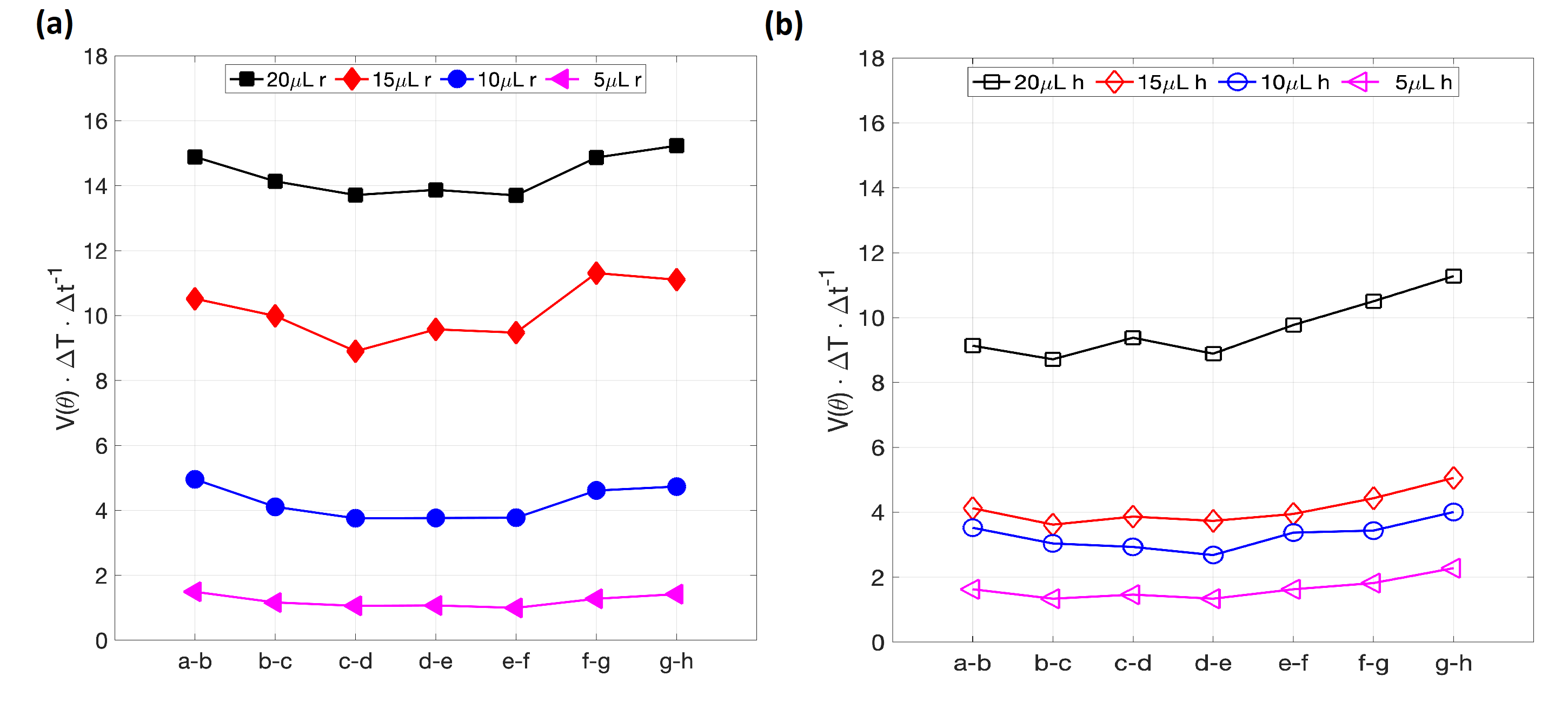}
    \caption{Volumetric thermal change for the (a) Room temperature drops and for the (b) hot drops.}
    \label{fig:VT}
\end{figure}

The curves resulting from the calculation presented in equation \eqref{eq:VT} are shown in Figures \ref{fig:VT}.a and \ref{fig:VT}.b. The U shape of the curve corresponds to the expected behaviour of the quotient $\frac{P_{PC}}{\rho c_d}$, that reflects the anomalous behavior of the density of water, which is maximum near $4\,$ºC, temperature reached between moments a-b and b-c. This result demonstrates that the contact angles obtained and the change observed in the shape of the drop are consistent with the known thermal characteristics of water.

\section{Discussion and conclusions}

The Mpemba effect, cold water freezes slower than hot water, is a challenging phenomenon that opens the opportunity to discuss the physical properties of water. 

Our research employs optical techniques like thermography to examine the Mpemba effect on droplets without affecting the temperature of cooling drops. We scrutinize alterations in the freezing process, selecting specific behavior changes on time and contact angles relative to temperature variations through a straightforward experimental setup. Our study used two non-invasive techniques: thermography and image analysis in the visible range.

It has been observed that drops at room temperature and hot drops exhibit different cooling dynamics. While many articles analyze the changes in contact angle and drop height during freezing, only a few studies have been conducted on the cooling process before recalescence. When we analyzed the energy involved in the freezing phenomena, we found no apparent difference between the hot and cold drops when the volumes are small; however, the higher amount of energy required for the room-temperature drop to freeze is more evident when the volume of the drop increases.

Using the contact angles, we obtained the volume of water droplets. We verified it against a characteristic curve showing the anomalous water density changes with temperature close to the freezing point.

We did not comment on the inverse thermal process, from the frozen drop to the melting point. However, it will be exciting to have an insight into it since, according to our observations, the behavior between drops at room and hot temperature is different, which implies that the water drop possesses a molecular memory. This study would be an interesting extension of this analysis for heat transfer laboratories at undergraduate-level.

Finally, we conclude by emphasizing that, inspired by the Mpemba effect, the study of the freezing of water droplets is crucial for various fields. Understanding the freezing process of water droplets is essential in cryopreservation, to improve techniques for preserving biological materials at low temperatures. Insights into phenomena like recalescence can enhance cryopreservation methods' efficiency and success rates by optimizing freezing protocols \cite{jingru,meng}. In meteorology, studying the freezing of water droplets aids in predicting weather patterns and understanding cloud formation processes \cite{jingru}. The freezing behavior of water droplets is significant in materials science for applications like ice-air jet technology and surface engineering for ice adhesion prevention coatings \cite{graeber}. These hot topics of applications motivate us to learn different, non-invasive techniques in the analysis of fluids. Also, interpreting our experiments requires considering various variables and other explanations for phenomena such as recalescence or water anomalous density behavior near 4 ºC. These are examples of the scientific methodology students need to understand the physics of actual and quotidian phenomena by collecting data and drawing conclusions.

%\begin{backmatter}
\subsection*{Acknowledges}
 A.B.O. acknowledges support from ``Investigadores por México'' CONAHCyT. E.H.F acknowledges support from ``Programa Ayudante de Investigador'' CONAHCyT. We thank Belem Patricia Falcon Varela and Julio Cesar Landa López for their motivational discussions on the Mpemba topic.
%\end{backmatter}

\bigskip
% Bibliography
%\bibliographystyle{unsrt}
%\bibliography{bibliography}

%\bibliography{bibliography}
\printbibliography
%\bibliographyfullrefs{bibliography}
\end{document}